\documentclass[aps,prl,amsmath,twocolumn,amssymb,superscriptaddress]{revtex4-1}

\usepackage{amssymb,amsfonts,amsmath}

\usepackage{color}
\usepackage[apple mac]{inputenc}
\usepackage[english]{babel}
\usepackage{times}
\usepackage{latexsym}
\usepackage{fancyhdr}
\usepackage{verbatim}
\usepackage{graphicx}
\usepackage{epsf}
\usepackage{subfigure}
\usepackage{bm}
\usepackage{epstopdf}\definecolor{Nathanblue}{rgb}{0.,0.24,0.51}
\newcommand{\blue}{\color{Nathanblue}}

\definecolor{orange}{rgb}{0.96,0.24,0.00}

\usepackage{mathtools}

\def\be{\begin{equation}}
\def\ee{\end{equation}}

\renewcommand{\i}{\mathrm{i}}

\newcommand{\ket}[1]{\ensuremath{\left| #1 \right\rangle}}
\newcommand{\bra}[1]{\ensuremath{\left\langle #1 \right|}}

\begin{document}

\title{{\blue Detecting fractional Chern insulators through circular dichroism}}

\author{C. Repellin}
\email[]{repellin@mit.edu}
\affiliation{Department of Physics, Massachusetts Institute of Technology, Cambridge, MA 02139, USA}
\author{N. Goldman}
\email[]{ngoldman@ulb.ac.be}
\affiliation{CENOLI,
Universit\'e Libre de Bruxelles, CP 231, Campus Plaine, B-1050 Brussels, Belgium}

\begin{abstract}
Great efforts are currently devoted to the engineering of topological Bloch bands in ultracold atomic gases. Recent achievements in this direction, together with the possibility of tuning inter-particle interactions, suggest that strongly-correlated states reminiscent of fractional quantum Hall (FQH) liquids could soon be generated in these systems. In this experimental framework, where transport measurements are limited, identifying unambiguous signatures of FQH-type states constitutes a challenge on its own. Here, we demonstrate that the fractional nature of the quantized Hall conductance, a fundamental characteristic of FQH states, could be detected in ultracold gases through a circular-dichroic measurement, namely, by monitoring the energy absorbed by the atomic cloud upon a circular drive. We validate this approach by comparing the circular-dichroic signal to the many-body Chern number, and discuss how such measurements could be performed to distinguish FQH-type states from competing states. Our scheme offers a practical tool for the detection of topologically-ordered states in quantum-engineered systems, with potential applications in solid state.
\end{abstract}

\date{\today}

\maketitle

\paragraph*{Introduction} Ultracold atomic gases constitute a promising platform for the study of strongly-correlated topological states~\cite{Cooper_review,Goldman_topology,Cooper_topology}, such as fractional Chern insulators (FCIs)~\cite{fractional_Chern_insulators, Bergholtz_fci_review}. Conceptually, these topological phases could be engineered in optical-lattice experiments, through the implementation of synthetic gauge fields~\cite{Dalibard_review,Goldman_review}. While such configurations have been realized~\cite{Aidelsburger2011,Aidelsburger2013,Miyake2013,Struck2013,Jotzu2014,Aidelsburger2015,Kennedy2015,Wu_2016,Huang2016,Tai2017,Flaschner2018,Tarnowski2018,Asteria2018}, so far, topological properties have only been identified in the non-interacting regime~\cite{Aidelsburger2015,Wu_2016,Flaschner2018,Tarnowski2018,Asteria2018}. 

In parallel to the development of schemes that would allow for the stabilization of strongly-correlated topological states in cold atoms~\cite{Cooper_review,Goldman_topology,Cooper_topology,Cooper-Dalibard2013,motruk-PhysRevB.96.165107,repellin-PhysRevB.96.161111,Reitter2017,Nager2018,Boulier2018,MesserHeating2018}, an open question  still remains:~Are there unambiguous probes for topological order that are applicable to interacting atomic systems? In the case of FCIs, the fractional nature of the quantized Hall conductance constitutes a striking manifestation of the underlying topological order~\cite{Yoshioka,fractional_Chern_insulators, Bergholtz_fci_review}. However, a direct access to the Hall conductance is challenging in optical-lattice setups; see Ref.~\cite{Lebrat2018} for conductance measurements in a one-dimensional optical lattice and Ref.~\cite{Salerno2018} for a possible extension to 2D. Alternative probing schemes have been proposed, such as the detection of edge excitations~\cite{Cazalilla_edge,Liu2010,Stanescu2010,Goldman2012_edge,Goldman2013_edge,Goldman2016_edge,Irsigler2018} and anyonic statistics~\cite{umucalilar2018time}, adiabatic pumps~\cite{Taddia2017,Wang2018,Raciunas2018}, and interferometric protocols involving mobile impurities bound to quasiparticles~\cite{Grusdt2016}.

\begin{figure}[h!]
\includegraphics[width = 0.96\linewidth]{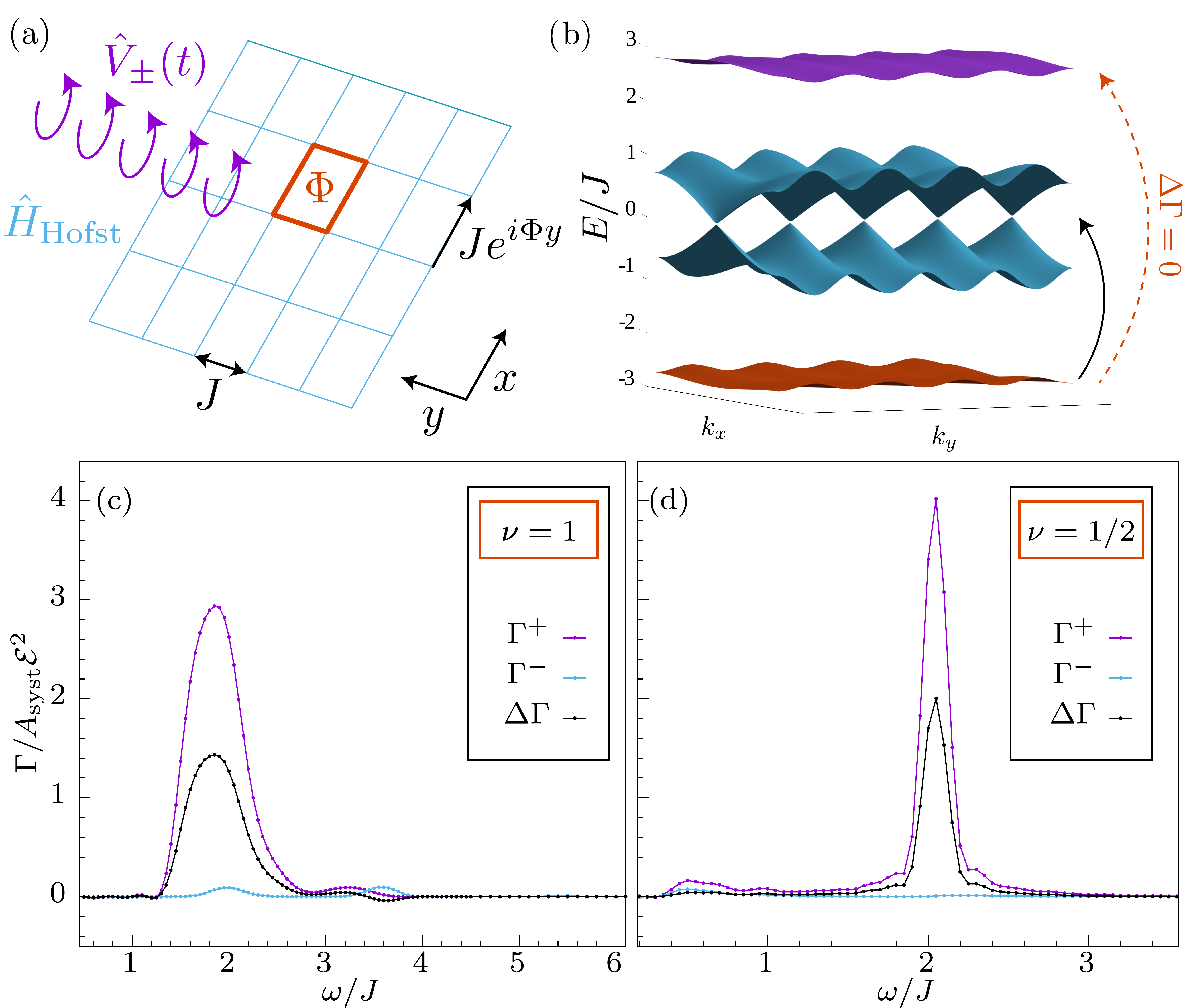}
\caption{Circular dichroism in the Harper-Hofstadter model. (a) Hofstadter model with flux $\Phi$ per plaquette, subjected to a circular drive $\hat{V}_{\pm}(t)$. (b) Single-particle spectrum for $\Phi\!=\!\pi/2$. The system is prepared in the ground state, such that only the lowest band is occupied (with fermions or bosons) at filling fraction $\nu$. The drive then induces transitions towards excited states at a rate $\Gamma_{\pm}(\omega)$. (c)~In the case of non-interacting fermions at $\nu\!=\!1$, the rates $\Gamma_{\pm}(\omega)$ describe inter-band transitions; for symmetry reasons, the differential response $\Delta \Gamma\!=\!(\Gamma_+\!-\!\Gamma_-)/2$ associated with the highest band is zero. (d) In the case of hard-core bosons at $\nu\!=\!1/2$, where the initial state is a Laughlin-type FCI, inter-band transitions still provide the largest contribution to the dichroic signal. The integral $\int \Delta \Gamma \text{d}\omega$ yields a quantized response that reveals topological order [Eq.~\eqref{DIR_fractional}].}
\label{fig 1}
\end{figure}

In this Letter, we demonstrate that the topological order of an atomic FCI could be  detected through circular-dichroic measurements, i.e.~by monitoring excitation rates upon circular shaking of the underlying 2D optical lattice~\cite{Tran2017,Tran2018,Asteria2018}.  Indeed, comparing the excitation rates associated with drives of opposite orientations ($\pm$), $\Delta \Gamma \!=\! (\Gamma_+-\Gamma_-)/2$, provides an indirect measurement of the imaginary part of the transverse optical conductivity~\cite{Tran2017,Bennett}
\be
\sigma_I^{xy} (\omega) = \hbar \omega \Delta \Gamma (\omega)/4 A_{\text{syst}}\mathcal{E}^2 , \label{optical_conductivity}
\ee
where $\mathcal{E}$ and $\omega$ are the amplitude and frequency of the circular drive, respectively, and $A_{\text{syst}}$ denotes the area of the 2D lattice. Then, combining Eq.~\eqref{optical_conductivity} with the Kramers-Kronig relations~\cite{Bennett,Hu1989,Souza}, one obtains a simple relation between the differential integrated rate (DIR), $\Delta \Gamma^{\text{int}}\!=\!\int_0^{\infty} \Delta \Gamma (\omega) \text{d} \omega$, and the Hall conductance $\sigma_{\text{H}}$ of the probed system~\cite{Tran2017,Tran2018}
\be
\Delta \Gamma^{\text{int}}/A_{\text{syst}} = (2 \pi \mathcal{E}^2/\hbar) \sigma_{\text{H}}.\label{DIR_Hall}
\ee
When applied to a (non-interacting) Chern insulator~\cite{bernevig2013topological}, characterized by the Chern number $\nu_{\text{Ch}} \!\in\! \mathbb{Z}$, Eq.~\eqref{DIR_Hall} gives rise to a quantization law for the DIR:~$\Delta \Gamma^{\text{int}}/A_{\text{syst}} \!=\! (\mathcal{E}/\hbar)^2 \nu_{\text{Ch}}$; we set the elementary charge $e\!=\!1$ to equally treat systems of charged and neutral gases~\cite{Dalibard_review,Goldman_review,Lebrat2018}. This quantized circular dichroism was recently measured in a non-interacting Fermi gas~\cite{Asteria2018}; see also Refs.~\cite{Anderson2017,brown2018bad} for indirect conductivity measurements in topologically-trivial Fermi gases. 

The relation in Eq.~\eqref{DIR_Hall} is universal in that it only relies on the causality of response functions~\cite{Bennett,Hu1989} and time-dependent perturbation theory~\cite{Bennett,Tran2017,Tran2018}. In particular, it applies to (strongly) interacting systems. In the case of a FCI, the Hall conductance is quantized in terms of a many-body Chern number $\sigma_{\text{H}}\!=\!(1/h) \nu_{\text{Ch}}^{\text{MB}}$, whose fractional value ($\nu_{\text{Ch}}^{\text{MB}} \!\in\! \mathbb{Q}$) is related to topological order~\cite{NiuThouless}. According to Eq.~\eqref{DIR_Hall}, excitation-rate measurements upon shaking could be used as a practical probe for the topological order of atomic FCIs,
\be
\Delta \Gamma^{\text{int}}/A_{\text{syst}} = (\mathcal{E}/\hbar)^2 \nu_{\text{Ch}}^{\text{MB}}, \qquad \nu_{\text{Ch}}^{\text{MB}} \!\in\! \mathbb{Q} . \label{DIR_fractional}
\ee


Here, we explore the validity and applicability of the relation in Eq.~\eqref{DIR_fractional} by performing exact-diagonalization studies on a realistic cold-atom model hosting FCI states:~the bosonic Harper-Hofstadter model~\cite{Aidelsburger2013,Miyake2013,Aidelsburger2015,Tai2017} with strong on-site interactions. By simulating the full time-evolution of this interacting system under the action of a circular drive, we demonstrate that the resulting excitation rates are indeed constrained by the topological order of the underlying phase [Eq.~\eqref{DIR_fractional}]. Our results indicate that the experimental protocol implemented in Ref.~\cite{Asteria2018} constitutes a realistic scheme by which topological order can be identified in strongly-interacting atomic systems.

\paragraph{The circular-dichroic measurement} 
We consider a generic many-body system described by a Hamiltonian $\hat{H}_0$, which is further subjected to a circular drive, 
\begin{equation}
\hat{H}_{\pm , \omega}\left( t \right) = \hat{H}_0 + 2\mathcal{E} \left[\cos(\omega t)\hat{x} \pm \sin(\omega t)\hat{y} \right] ,
\label{eq: drive}
\end{equation}
where  $\omega$,  $\mathcal E$ and $\pm$ refer to the frequency, amplitude and orientation of the drive, respectively; $\hat{x}, \hat{y}$ are position operators. 
We assume that the system is initially prepared in the ground-state $\ket{\Psi_0}$ of $\hat{H}_0$, with energy $E_0$, and we analyze the probability of finding the system in an excited state at a given time, $\mathcal{P}_{\pm , \omega}(t) \!=\!1\!-\!|\left<\Psi_0|\Psi(t)\right>|^2$, for a given configuration of the drive. The circular-dichroic signal $\Delta \Gamma (\omega)$ entering Eq.~\eqref{optical_conductivity} is then obtained by evaluating the excitation rates $\Gamma_{\pm}(\omega)\!=\!\mathcal{P}_{\pm,\omega}(t) / t$; the power absorbed upon driving reads $P_{\pm}(\omega)\!=\!\hbar \omega \Gamma_{\pm}(\omega)$. Figure~\ref{fig 1} displays two illustrative spectra $\Gamma_{\pm}(\omega)$, corresponding to a Chern insulator of non-interacting fermions and a FCI of strongly-interacting bosons.

\begin{figure}
\includegraphics[width=0.98\linewidth]{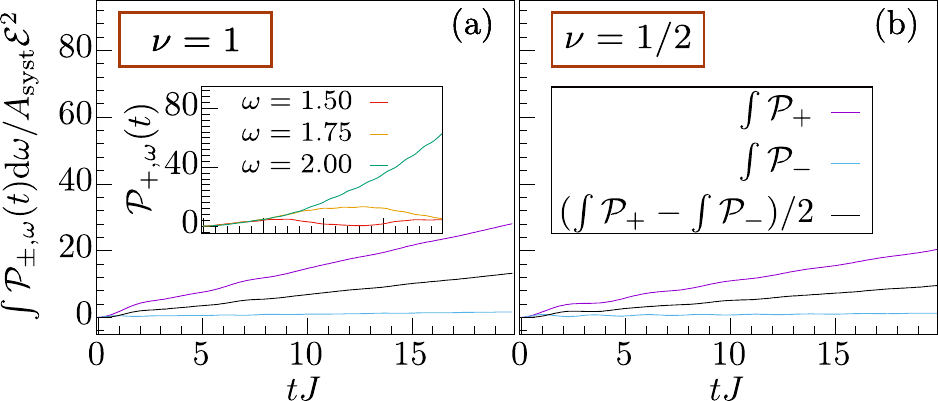}
\caption{Probability of scattering into an excited state upon circular driving, integrated over the drive frequency. (a) System of $8$ non-interacting fermions at filling fraction $\nu\!=\!1$. The inset shows the probabilities $\mathcal{P}_{+ , \omega}(t)$ for individual values of $\omega$ in units of $A_{\text{syst}}\mathcal{E}^2$. (b) System of $7$ hard-core bosons at filling fraction $\nu\!=\!1/2$. These results indicate the typical time-scale $(t\!\sim\!10\hbar/J)$ over which measurements should be performed so as to extract $\Delta \Gamma^{\text{int}}$ with accuracy.}
\label{fig 2}
\end{figure}

In practice, the extraction of the excitation rates $\Gamma_{\pm}(\omega)$ entering Eqs.~\eqref{optical_conductivity}-\eqref{DIR_Hall} can be complicated by finite-size effects, which are intrinsically present in numerical studies; importantly, similar effects could also alter experimental realizations involving small atomic ensembles~\cite{Tai2017}. In order to appreciate this aspect, let us recall that the results in Eqs.~\eqref{optical_conductivity}-\eqref{DIR_Hall} stem from time-dependent perturbation theory~\cite{Cohen_book,Tran2017}, which expresses the (constant) excitation rates as [henceforth, $\hbar\!=\!1$]
\begin{align}
&\Gamma_{\pm} (\omega)= 2 \pi \mathcal{E}^2 \sum_{n\ne 0} \vert \langle \Psi_n \vert \hat x \pm i \hat y \vert \Psi_0 \rangle \vert^2 \delta^{(t)} (E_n - E_0 - \omega) , \notag\\
&\delta^{(t)}(E)\!=\!(2/\pi t) \sin^2 (E t/2)/E^2 \, \xrightarrow[]{t \rightarrow \infty} \delta(E), \label{rate_gen_real}
\end{align}
where  $\vert \Psi_n \rangle$ denotes the excited states with energy $E_n$. We point out that Eq.~\eqref{rate_gen_real} relies on two assumptions~\cite{Cohen_book}:~(a) the rotating-wave approximation is satisfied, $t\!\gg\!1/ \omega$; and (b) the observation time is small compared to the Rabi periods $t\!\ll\!1/ \mathcal{E} \vert \langle \Psi_n \vert \hat x \pm i \hat y \vert \Psi_0 \rangle \vert$, associated with all possible transitions. For small system sizes, the complete time-evolution associated with Eq.~\eqref{eq: drive} is affected by the discrete nature of the spectrum $E_n$, which results in residual oscillations in the excitation probability $\mathcal{P}_{\pm , \omega}(t)$; see inset of Fig.~\ref{fig 2}a. This often prevents a precise evaluation of the rates $\Gamma_{\pm} (\omega)$. Since we are ultimately interested in the integrated rates [Eq.~\eqref{DIR_Hall}], we get rid of these pathological oscillations by integrating $\mathcal{P}_{\pm , \omega}(t)$ over all relevant frequencies. This leads to the quasi-linear behavior shown in Fig.~\ref{fig 2}, which allows for an efficient evaluation of the DIR, $\Delta \Gamma^{\text{int}}$, entering Eqs.~\eqref{DIR_Hall}-\eqref{DIR_fractional}.

We now turn to the right-hand side of Eq.~\eqref{DIR_fractional}:~the expression for the many-body Chern number $\nu_{\text{Ch}}^{\text{MB}}$. In the single-particle context, the Chern number is defined as the integral of the Berry curvature $F$ over the first Brillouin zone~\cite{qi2011topological,bernevig2013topological}. The many-body counterpart of the Berry curvature can be obtained by considering twisted boundary conditions~\cite{NiuThouless}, with angles $\mathbf{\theta}_{x,y}$, and by considering partial derivatives of the many-body ground-state over these variables:~$F(\mathbf{\theta}_x,\mathbf{\theta}_y) = \partial_{\theta_x}A_y - \partial_{\theta_y}A_x$, 
where $A_\mu\!=\!(1/d)\text{Tr}\left(\Psi_i\partial_{\theta_\mu}\Psi_j\right) $ is the generalized Berry connection, and where the trace $(1/d)\text{Tr}$ averages over the manifold spanned by the $d$-degenerate many-body ground states $\Psi_i$. Integrating $F(\mathbf{\theta}_x,\mathbf{\theta}_y)$ over the angles $\theta_{x,y}$ yields a quantized number~\cite{NiuThouless,hafezi-PhysRevA.76.023613}, which traditionally defines the many-body Chern number $\nu_{\text{Ch}}^{\text{MB}}$. Here, we instead consider the nonintegrated many-body Chern number, which is defined as
\begin{equation}
\bar \nu_{\text{Ch}}^{\text{MB}} \equiv -2\pi i F(\theta_x \!=\! 0 , \theta_y \!=\! 0 ).
\label{eq: nu_ch}
\end{equation}
This quantity allows for a faithful evaluation of the Hall conductance entering Eqs.~\eqref{DIR_Hall}-\eqref{DIR_fractional}, as can be obtained through the Kubo formula~\cite{NiuThouless,kudo2018many,Watanabe2018}. We stress that $\bar \nu_{\text{Ch}}^{\text{MB}}$ is not quantized, however it converges towards $\nu_{\text{Ch}}^{\text{MB}}$ in the thermodynamic limit~\cite{Hastings2015, koma-2015arXiv150401243K, Bachmann2018}. As a byproduct of our analysis, we will hereby provide the scaling of $\bar \nu_{\text{Ch}}^{\text{MB}}$ with respect to the system size, in the FCI regime.

\paragraph{Microscopic model} We explore the relation in Eq.~\eqref{DIR_fractional} by studying the bosonic Harper-Hofstadter Hamiltonian, 
\begin{align}
\hat{H}_0=&-J\left(\sum_{m, n} \hat{a}^\dagger_{m, n+1} \hat{a}_{m,n} + e^{i\frac{\pi n}{2}}\hat{a}^{\dagger}_{m+1, n}\hat{a}_{m,n} + \text{h.c.} \right) \notag \\
&+(U/2) \sum_m \hat{a}^\dagger_{m, n}\hat{a}_{m,n} (\hat{a}^\dagger_{m, n}\hat{a}_{m,n}-1),
\label{eq: time independent H}
\end{align}
which describes the hopping of bosons over a square lattice, with tunneling amplitude $J$, in the presence of a flux $\Phi$ per plaquette~\cite{hofstadter1976energy} and on-site interactions $U$. Here, we set the flux $\Phi\!=\!\pi / 2$, as was recently realized in cold-atom experiments~\cite{Aidelsburger2013,Aidelsburger2015,Tai2017}, and impose periodic boundary conditions~\cite{footnote open systems}. 

We consider the ground-state properties of this model for a fixed filling fraction $\nu\!=\!\frac{N}{N_x N_y}$, where $N$ denotes the total number of particles and $N_x \times N_y$ is the number of magnetic unit cells. In the present case, where $\Phi\!=\!\pi / 2$ and the Chern number of the lowest band is $\nu_{\text{Ch}}\!=\!1$, we are specifically interested in the filling $\nu\!=\!1/2$ of bosons in the presence of strong repulsive interactions. For the sake of simplicity and numerical performance, we suppose that the interaction is much larger than the band gap and we include this effect in the form of a hard-core constraint for the bosons. The ground state of this system was found to be a FCI~\cite{sorensen-PhysRevLett.94.086803, hafezi-PhysRevA.76.023613, repellin-PhysRevB.96.161111, he-PhysRevB.96.201103, motruk-PhysRevB.96.165107}, akin to the Laughlin state exhibiting the FQH effect~\cite{Yoshioka}. Its most striking signatures of topological order include a non-trivial (twofold) degeneracy and a fractional many-body Chern number, $\nu_{\text{Ch}}^{\text{MB}}\!=\!1/2$.

In order to probe the topological nature of the ground state ($\nu_{\text{Ch}}^{\text{MB}}\!=\!1/2$) through circular dichroism [Eq.~\eqref{DIR_fractional}], we analyze its response to a circular perturbation [Eq.~\eqref{eq: drive}]. This study is performed in a moving frame where the total Hamiltonian in Eq.~\eqref{eq: drive} is translationally invariant, as generated by the operator $\hat{R}_{\pm}(t)\! =\! \exp\left \{i (2 \mathcal{E} /\omega) \left [\hat{x}\sin(\omega t) \mp \hat{y} \cos(\omega t) \right ]\right \}$.
Note that the (on-site) interactions in Eq.~\eqref{eq: time independent H} are not affected by this change of frame. The resulting time-dependent Hamiltonian reads
\begin{align}
\hat{{\cal H}}_{\pm , \omega}(t)  =  -J\big(&\sum_{m, n}e^{ \mp i \frac{2\mathcal{E}}{\omega}\cos(\omega t)} \hat{a}^\dagger_{m, n+1} \hat{a}_{m,n} \label{eq: H(t)} \\ 
                    &  + e^{i\frac{\pi n}{2}+i \frac{2\mathcal{E}}{\omega} \sin(\omega t)}\hat{a}^{\dagger}_{m+1, n}\hat{a}_{m,n} + \text{h.c.} \big),\nonumber 
\end{align}
where the hard-core constraint is henceforth implicit, except otherwise stated.

\paragraph{Numerical results}
\begin{figure}
\includegraphics[width=0.98\linewidth]{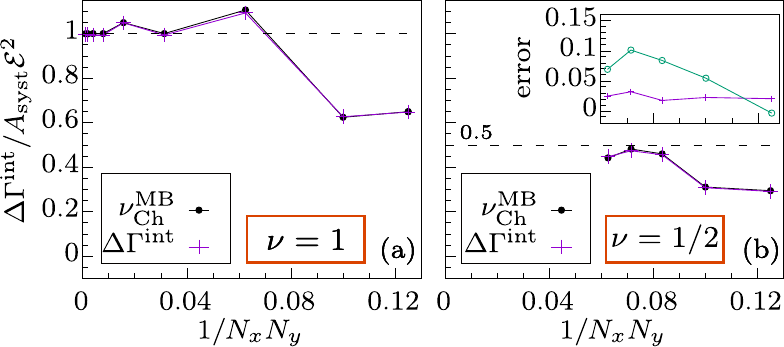}
\caption{Dichroic signal and nonintegrated many-body Chern number [Eq.~\eqref{eq: nu_ch}] as a function of the system size. (a) In the case of non-interacting fermions at filling fraction $\nu\!=\!1$, both the dichroic signal and $\bar \nu_{\text{Ch}}^{\text{MB}}$ converge towards $\nu_{\text{Ch}}\!=\!1$ in the thermodynamic limit. (b) In the case of hard-core bosons at $\nu\!=\!1/2$, the results are limited to smaller systems, but show that these quantities both approach the quantized value $\nu_{\text{Ch}}^{\text{MB}}\!=\!1/2$. In all cases, the dichroic signal and $\bar \nu_{\text{Ch}}^{\text{MB}}$ perfectly overlap for all sizes; see the relative error in inset (purple curve). Restricting the evaluation of the DIR to inter-band transitions only leads to the small relative error shown by the green curve (inset).}
\label{fig 3}
\end{figure}
Our numerical protocol starts with the calculation of the ground state associated with the Hamiltonian in Eq.~\eqref{eq: time independent H} using exact diagonalization.  We subsequently turn on the periodic drive and perform a stroboscopic time-evolution of the system, by repeatedly applying the time-evolution operator associated with Eq.~\eqref{eq: H(t)} over small time steps. 
Interacting-boson calculations are performed for various finite-size clusters, made of up to $64$ lattice sites (or $N_xN_y\!=\!16$ magnetic unit cells).

As a benchmark, we first apply the protocol to the case of non-interacting fermions at filling $\nu\!=\!1$, hence forming a Chern insulator~\cite{qi2011topological,bernevig2013topological}. The circular dichroism of a Chern insulator was previously analyzed based on a two-band  model~\cite{Tran2017,Asteria2018}. We hereby extend these results to the multi-band Hofstadter model in the first line of Eq.~\eqref{eq: time independent H}, whose energy spectrum is illustrated in Fig.~\ref{fig 1}b. Figure~\ref{fig 1}c shows the numerical excitation rates $\Gamma_{\pm}(\omega)$, for a system formed of $32\!\times\!64$ magnetic unit cells. An intense absorption peak is clearly identified around $\omega\!=\! 2J$, which corresponds to excitations to the middle band [Fig.~\ref{fig 1}b]; this peak is associated with a large dichroic signal $\Delta \Gamma$. In contrast, transitions to the highest band do not contribute to the differential rate $\Delta \Gamma$, as can be seen from the absence of any signal above $\omega\!=\!4J$ in Fig.~\ref{fig 1}c. This phenomenon is due to a special symmetry between the lowest and highest bands of the Harper-Hofstadter model~\cite{Aidelsburger2015}, which is found for any flux $\Phi$; see~\cite{SupMat} for details. Using the methods described above, we finely extract the DIR $\Delta \Gamma^{\text{int}}$, and compare it to the finite-size Chern number $\bar \nu_{\text{Ch}}^{\text{MB}}$ defined in Eq.~\eqref{eq: nu_ch}:~as shown in Fig.~\ref{fig 3}a, the dichroic signal $\Delta\Gamma^{\text{int}}/(A_{\text{syst}} \mathcal{E}^2)$ perfectly overlaps with $\bar \nu_{\text{Ch}}^{\text{MB}}$, and these quantities both converge towards the quantized value $\nu_{\text{Ch}}\!=\!1$, in perfect agreement with Eq.~\eqref{DIR_Hall}.

We now turn to the case of strongly interacting (hard-core) bosons at filling fraction $\nu\!=\!1/2$. The ground state of this system has an (approximate) twofold degeneracy on the torus, which is characteristic of a Laughlin-type FCI at this filling factor. Similarly to the definition of $\bar\nu_{\text{Ch}}^{\text{MB}}$ in Eq.~\eqref{eq: nu_ch}, we evaluate the dichroic signal by performing an average over the ground-state manifold; we found that the contributions of the two relevant states can be estimated individually, any residual mixing between them contributing negligibly to the dichroic signal. In order to explore the law in Eq.~\eqref{DIR_fractional}, we analyzed the real-time depletion of the ground-state manifold upon the circular drive; we extracted $\Delta\Gamma^{\text{int}}$ from the quasi-linear curves obtained by integrating the excitation probabilities over $\omega$ [Fig.~\ref{fig 2}b]. 
As a central result of our studies, we demonstrate in Fig.~\ref{fig 3} that the resulting dichroic signal $\Delta\Gamma^{\text{int}}/(A_{\text{syst}} \mathcal{E}^2)$ is in excellent agreement with the finite-size Chern number $\bar \nu_{\text{Ch}}^{\text{MB}}$ in Eq.~\eqref{eq: nu_ch}, for all system sizes.  These results provide good indications that the value $\Delta\Gamma^{\text{int}}/(A_{\text{syst}} \mathcal{E}^2)\approx\bar\nu_{\text{Ch}}^{\text{MB}}\!\rightarrow\!\nu_{\text{Ch}}^{\text{MB}}\!=\!1/2$ will be reached in the thermodynamic limit, as expected for a Laughlin-type FCI.

\begin{figure}
\includegraphics[width=0.99\linewidth]{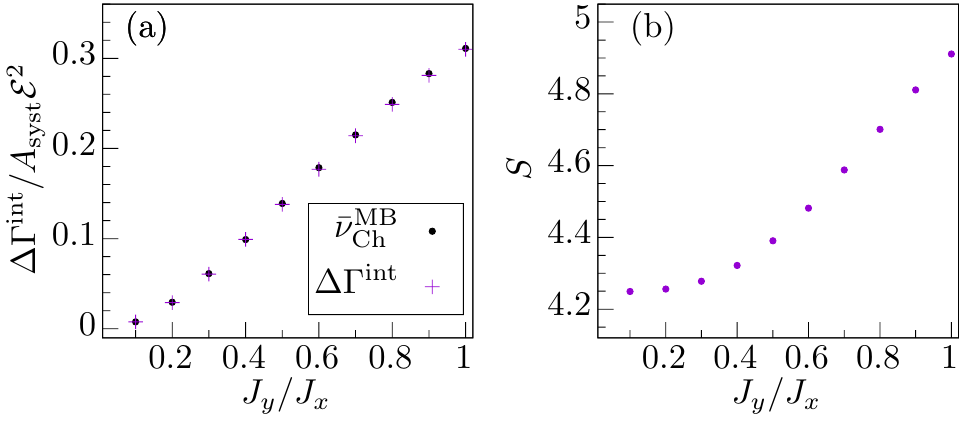}
\caption{Tracking the transition from a CDW to a FCI: (a) The dichroic signal, and (b) the entanglement entropy upon particle cut, as one modifies the ratio of tunneling amplitudes $J_y/J_x$ in the lattice; here, $N\!=\!7$ bosons and $N_xN_y\!=\!14$ magnetic cells ($\nu\!=\!1/2$). The dichroic signal distinguishes the CDW ($J_y/J_x\!\ll \!1$, $\Delta\Gamma^{\text{int}}/A_{\text{syst}}\mathcal{E}^2\!\simeq\!0$) from the FCI ($J_y/J_x\!\simeq\!1$, $\Delta\Gamma^{\text{int}}/A_{\text{syst}}\mathcal{E}^2\!\simeq\!1/2$), and it perfectly overlaps with $\bar \nu_{\text{Ch}}^{\text{MB}}$ throughout the transition.}
\label{fig 4}
\end{figure}


While excitations are simple inter-band transitions in the non-interacting Chern-insulator case, they may be richer in the FCI due to its intrinsic many-body nature. In this regard, the absorption spectrum $\Delta \Gamma (\omega)$ in Fig.~\ref{fig 1}d, which characterizes the FCI and its excitations, delivers interesting insights.  In our perturbative framework, the rates in Eq.~\eqref{rate_gen_real} are associated with single absorption processes, hence, the probed excitations $\Psi_n$  are limited to single-particle excitations of the FCI. This explains the limited range ($\omega_{\text{min}}\!=\!0.25J\!<\!\omega\!<\!\omega_{\text{max}}\!=\!3.5J$) over which the dichroic signal is observed, which sets a small upper bound $\omega_{\text{max}}$ for the frequency integration window. The ``activation" frequency $\omega_{\text{min}}$ is predictably close to the many-body gap $\Delta\!\simeq\!0.2J$~\cite{repellin-PhysRevB.96.161111, repellin-PhysRevB.90.045114}, which corresponds to the minimum of the magneto-roton mode~\cite{GMP}. The response associated with this collective mode, which was identified as a single-particle modulation of Laughlin-type states~\cite{GMP,repellin-PhysRevB.90.045114}, appears as a low-frequency peak in the absorption spectrum in Fig.~\ref{fig 1}d. Besides, the large absorption peak at $\omega\!\simeq\!2J$, which coincides with the main signal of Fig.~\ref{fig 1}c, suggests the predominance of inter-band transitions in the dichroic signal. To confirm this observation, we have isolated the exact contribution of inter-band transitions to $\Delta\Gamma^{\text{int}}$ by evaluating the dynamical repopulation of the bands upon  circular driving. The results are summarized in the inset of Fig.~\ref{fig 3}, which shows the relative error to the dichroic signal when restricting its evaluation to inter-band transitions only. 
For all $N$ considered, we find that the inter-band transitions constitute more than $90\%$ of the DIR ($100\%$ for the smallest system size, $N\!=\!4$).
This finding is of crucial experimental importance, as band repopulations are routinely measured in cold atoms through band-mapping techniques~\cite{Aidelsburger2015,Asteria2018}. As a byproduct of our study, we also obtain the transverse optical conductivity $\sigma_I^{xy}(\omega)$ of the FCI, by inserting the signal $\Delta \Gamma (\omega)$ into Eq.~\eqref{optical_conductivity}; see~\cite{SupMat}.

\paragraph{Phase transition} 

In the thermodynamic limit, the fractional value of the Chern number $\nu_{\text{Ch}}^{\text{MB}}$ is an unambiguous signature of topological order~\cite{NiuThouless}. However, the actual values of $\bar \nu_{\text{Ch}}^{\text{MB}}\!\approx\!\Delta\Gamma^{\text{int}}/A_{\text{syst}}\mathcal{E}^2$ may significantly differ from the quantized $\nu_{\text{Ch}}^{\text{MB}}$ when restricting the analysis to small system sizes [Fig.~\ref{fig 3}]; this fact could alter the ability of the dichroic signal to distinguish topologically-ordered phases (FCIs) from competing phases in cold-atom experiments based on small atomic ensembles.
To address this issue, we analyze the behavior of these observables across a phase transition. We consider a situation where the tunneling amplitudes along the two directions of the lattice, $J_{x,y}$, can be tuned independently from $J_y/J_x\!=\!0$ (decoupled Luttinger liquids) to $J_y/J_x\!=\!1$:~at small enough ratio $J_y/J_x$, the ground state shares characteristics of a charge density wave (CDW) in the lowest Bloch band (as observed in the ``thin-torus" limit of FCIs~\cite{budich-PhysRevB.88.035139, bernevig-2012arXiv1204.5682B}); this phase transition was proposed as a method to prepare atomic FCIs~\cite{he-PhysRevB.96.201103}. Despite the absence of topological order in the CDW, some of its signatures are similar to those of a FCI, e.g.~a twofold degeneracy on the torus due to the spontaneous breaking of discrete translation symmetry. Our results show that the circular-dichroic signal distinguishes the CDW from FCI phases, even for very small systems [Fig.~\ref{fig 4}a]. These results are compared with entanglement signatures in Fig.~\ref{fig 4}b, which can also distinguish between these two competing phases~\cite{budich-PhysRevB.88.035139, bernevig-2012arXiv1204.5682B}, but are not directly accessible in experiments~\cite{footnote}. 

\paragraph{Concluding remarks} Several schemes have been proposed to generate atomic FCIs~\cite{sorensen-PhysRevLett.94.086803,Cooper-Dalibard2013,yao2013realizing,motruk-PhysRevB.96.165107,repellin-PhysRevB.96.161111,he-PhysRevB.96.201103,grusdt2014realization}. The circular-dichroic probe can be readily applied to any of these cold-atom configurations. Indeed, the protocol described in the previous paragraphs only requires two main ingredients: (a) the implementation of a circular perturbation [Eq.~\eqref{eq: drive}], which can be induced by shaking a 2D optical lattice~\cite{Struck2013,Jotzu2014,Asteria2018}; and (b) a probe for the excitation rates $\Gamma_{\pm}$. Since the DIR entering Eqs.~\eqref{DIR_Hall}-\eqref{DIR_fractional} has predominant contributions from inter-band transitions, the related rates can be simply evaluated using band-mapping techniques~\cite{Asteria2018}. When intra-band transitions are significant, a more sophisticated probe should be considered; e.g.~one could exploit pulsed magnetic field gradients~\cite{xu2013atomic,anderson2013magnetically,goldman2014periodically} to generate a circular drive $\hat V_{\pm}(t)\!\sim\!\left [\hat x \sigma_x \cos(\omega t) \pm \hat y \sigma_y \sin(\omega t) \right ]$, which simultaneously changes the internal state ($\sigma$) of the atoms, hence allowing for a complete read-out of the excited fraction through state-dependent imaging~\cite{Goldman2012_edge}. In principle, the probe in Eq.~\eqref{eq: drive} could also be considered as an alternative to transport measurements in solid state:~there materials would be irradiated by circularly-polarized light~\cite{Bennett} and the resulting DIR extracted from optical absorption or photoemission measurements~\cite{wang2013circular,kim2016chiral,schuler2017tracing}.

\begin{acknowledgments}
\paragraph*{Acknowledgments} We acknowledge fruitful discussions with N.R. Cooper, J. Dalibard, M. Dalmonte, F. Grusdt, A. Grushin, G. Jotzu, D.T. Tran, C. Weitenberg and P. Zoller. We also thank A. Sterdyniak and T. Yefsah for an earlier collaboration on a related topic. Work in Brussels is supported by the FRS-FNRS (Belgium) and the ERC Starting Grant TopoCold. C.R. is supported by the Marie Sklodowska-Curie program under EC Grant agreement 751859.
\end{acknowledgments}

\clearpage
\newpage

\begin{center} 
 {\blue {\Large Supplementary material}}
\end{center}

\subsection{Chiral symmetry and circular dichroism \\ in Hofstadter bands}

Analyzing the circular dichroism of the non-interacting Harper-Hofstadter model led to a surprising result:~As observed in Fig.1c (main text), the dichroic signal $\Delta \Gamma (\omega)$ only involves transitions from the lowest to the middle bands; in other words, the highest band of the model [Fig.1b; main text] does not contribute to this circular dichroic effect. In fact, this property is due to a special (chiral) symmetry that relates the lowest and highest bands of the Harper-Hofstadter model, and which is present for any value of the flux $\Phi$. It is the aim of this Appendix to provide the detailed proof of this assertion.

We start by briefly reviewing the chiral symmetry inherent to the Harper-Hofstadter model and we discuss how it affects the Berry curvature in the lowest band. In this Appendix, we consider the Harper-Hofstadter model with a generic flux $\Phi\!=\!2\pi(p/q)$, where $p,q$ are relative primes. The corresponding single-particle Hamiltonian can be written as~[79]
\begin{align}
\hat h_0&= -J \sum_{m,n} \vert m,n+1 \rangle \langle m,n \vert + e^{i \Phi n} \vert m + 1,n \rangle \langle m,n \vert + \text{h.c.} \notag\\
&=-J\sum_{{\bf k}} 2 \cos k_y \vert {\bf k}\rangle\langle {\bf k}\vert + \left (e^{i k_x} \vert k_x, k_y - \Phi \rangle\langle {\bf k}\vert + \text{h.c.} \right ),\notag
\end{align}
where we considered both the real-space and momentum-space representations. Here the state $\vert m,n \rangle$ represents an orbital localized at lattice site $(m,n)$ of the square lattice, whereas ${\bf k}\!=\!(k_x,k_y)$ denote the momenta. Diagonalizing this Hamiltonian yields $q$ Bloch bands~[60]; we will denote the corresponding eigenstates as $\vert n_{\bf k} \rangle$ and eigenvalues as $E_n ({\bf k})$, where $n\!=\!1, \dots, q$ is the band index.

\subsubsection{Chiral symmetry in the Hofstadter model}

To highlight the aforementioned chiral symmetry, we introduce a ``checkerboard" bipartition $A/B$ of the square lattice, such that the neighbors of all $A$ sites are $B$ sites and vice versa, and we define the following operator 
\begin{align}
\hat{\Pi} = \hat{P}_A - \hat{P}_B&= \sum_{m,n} e^{i \pi (m+n)} \vert m,n \rangle \langle m,n \vert \notag \\
&=\sum_{{\bf k}} \vert k_x - \pi, k_y -\pi \rangle\langle {\bf k}\vert , \label{pi_operator}
\end{align}
where $\hat{P}_A$ and $\hat{P}_B$ project onto the $A$ and $B$ sites, respectively. One readily verifies that the Harper-Hofstadter Hamiltonian $\hat{h}_0$ anticommutes with $\hat{\Pi}$, $\{ \hat{h}_0, \hat{\Pi}\}\!=\!0$, which leads to the ``particle-hole" symmetry of the whole spectrum:~Any eigenstate $\vert n_{\bf k} \rangle$ of $\hat{h}_0$ with energy $E$ has a partner eigenstate $\hat{\Pi}\vert n_{\bf k} \rangle$ with energy $-E$. 

In order to make this statement more specific, in particular, with respect to the quasi-momentum $\mathbf{k}$ of the eigenstates, let us examine the commutation relations involving $\hat{\Pi}$ and the magnetic-translation operators along the $x$ and $y$ directions, which we denote $\hat{T}_{x,y}$, respectively. If the magnetic unit cell comprises an even number of lattice sites ($q$ even), the action of $\hat{T}_{x,y}$ leaves the bipartition unchanged, such that $[\hat{\Pi}, \hat{T}_{x,y}]\!=\! 0$. Conversely, if $q$ is odd, $\hat{T}_{x,y}$ exchanges the positions of $A$ and $B$, such that $[\hat{\Pi}, \hat{T}_{x,y}]\!\neq\!0$, however, the operators satisfy $[\hat{\Pi}, \hat{T}_{x,y}^2]\!=\!0$. Then, the following symmetry property between the wave functions of the highest ($n\!=\!q$) and lowest ($n\!=\!1$) bands follows
\begin{align}
&\hat{\Pi} \ket{1_{\mathbf{k}}}  = \ket{q_{\mathbf{k}}} & \text{ for }q \text{ even} \label{q even}, \\
&\hat{\Pi} \ket{1_{\mathbf{k}}}  = \ket{q_{\mathbf{k+\boldsymbol{\pi}}}} & \text{ for } q \text{ odd} \label{q odd},
\end{align}
where $\mathbf{k+\boldsymbol{\pi}}$ corresponds to a $\pi$ shift of momentum along the $x$ and $y$ directions [Eq.~\eqref{pi_operator}]; an alternative demonstration of this result is provided in the Supplementary Information of Ref.~[13].

\subsubsection{Chiral symmetry and the Berry curvature}

We now analyze how these symmetries affect the Berry curvature $\Omega_{xy}(\mathbf{k})$ in the lowest band ($n\!=\!1$). By definition, the latter can be expressed as a sum of contributions from all the other bands~[80]
\begin{align}
&\Omega_{xy}(\mathbf{k})  =  \sum_{n>1}  \Omega^n_{xy} (\mathbf{k}), \notag \\
&\Omega^n_{xy} (\mathbf{k})  =  \text{Im}\left[\frac{\bra{1_\mathbf{k}}\partial_{k_x}\hat{h}_0\ket{n_\mathbf{k}} \bra{n_\mathbf{k}} \partial_{k_y}\hat{h}_0 \ket{1_\mathbf{k}}}{[E_n(\mathbf{k}) - E_1(\mathbf{k})]^2}\right].
\label{eq: Berry curvature}
\end{align}

Let us first consider the case where $q$ is even (as in the main text, where $q\!=\!4$). In this case, the contribution of the highest band ($n\!=\!q$) to the Berry curvature in Eq.~\eqref{eq: Berry curvature} reads
\begin{align}
\Omega^q_{xy} (\mathbf{k})  &=  \text{Im}\left[\frac{\bra{1_\mathbf{k}}\partial_{k_x}\hat{h}_0\ket{q_\mathbf{k}} \bra{q_\mathbf{k}} \partial_{k_y}\hat{h}_0 \ket{1_\mathbf{k}}}{[E_q(\mathbf{k}) - E_1(\mathbf{k})]^2}\right] \notag\\
&\propto\text{Im}\left[\bra{1_\mathbf{k}}\partial_{k_x} (\hat{h}_0 \hat{\Pi})\ket{1_\mathbf{k}} \bra{1_\mathbf{k}} \partial_{k_y} (\hat{\Pi} \hat{h}_0) \ket{1_\mathbf{k}}\right] \notag \\
&=0 , \label{qzero}
\end{align}
where we first used the symmetry property in Eq.~\eqref{q even}, and then the fact that the diagonal matrix elements of any skew-hermitian operator are purely imaginary (we note that the product $\hat{h}_0 \hat{\Pi}$ is skew-hermitian since  $\{ \hat{h}_0, \hat{\Pi}\}\!=\!0$). Consequently, in the Harper-Hofstadter model with $q$ even, the highest band never contributes to the local Berry curvature $\Omega_{xy}(\mathbf{k})$ of the lowest band. 

In the case where $q$ is odd, we find a less restrictive condition
\begin{align}
\Omega^q_{xy} (\mathbf{k}) = -  \Omega^q_{xy} (\mathbf{k+\boldsymbol{\pi}}),
\label{qodd}
\end{align}
which can be obtained by observing that the matrix elements and energies entering Eq.~\eqref{eq: Berry curvature} satisfy
\begin{align}
&\bra{1_{\mathbf{k+\boldsymbol{\pi}}}}\partial_{k_x}\hat{h}_0\ket{q_{\mathbf{k+\boldsymbol{\pi}}}}  \bra{q_{\mathbf{k+\boldsymbol{\pi}}}} \partial_{k_y}\hat{h}_0 \ket{1_{\mathbf{k+\boldsymbol{\pi}}}}\notag \\
&\qquad =\left (\bra{1_\mathbf{k}}\partial_{k_x}\hat{h}_0\ket{q_\mathbf{k}} \bra{q_\mathbf{k}} \partial_{k_y}\hat{h}_0 \ket{1_\mathbf{k}} \right )^* , \label{matrix_elements} \\
&E_q(\mathbf{k}+\boldsymbol{\pi}) - E_1(\mathbf{k}+\boldsymbol{\pi}) = E_q(\mathbf{k}) - E_1(\mathbf{k}).\label{energies_relation}
\end{align}
We point out that the relations in Eqs.~\eqref{matrix_elements}-\eqref{energies_relation} result from the  chiral symmetry discussed in the previous paragraph. From Eqs.~\eqref{qzero}-\eqref{qodd}, we deduce that the highest band never contributes to the integrated Berry curvature (Chern number) of the lowest band, for any value of the flux ($q$ odd or even).

\subsubsection{Circular dichroism in Hofstadter bands}

These observations have a striking impact on the circular dichroism of a Chern insulator that is prepared in the lowest band of the Harper-Hofstadter model. To see this, we recall how the dichroic signal relates to the Berry curvature. Considering a completely filled lowest band, the dichroic signal can be expressed as~[43]
\begin{align}
\Delta \Gamma (\omega) =  4\pi \mathcal{E}^2 \sum_{\mathbf{k}} \sum_{n>1} &\text{Im}\left[\frac{\bra{1_\mathbf{k}}\partial_x\hat{h}_0\ket{n_\mathbf{k}} \bra{n_\mathbf{k}} \partial_y\hat{h}_0 \ket{1_\mathbf{k}}}{[E_n(\mathbf{k}) - E_1(\mathbf{k})]^2}\right] \notag \\ &\times \delta^{(t)} (E_n(\mathbf{k}) - E_1(\mathbf{k}) - \omega) , 
\end{align}
where the function $\delta^{(t)}(E)$ is defined in the main text [Eq.~(5)]. In the long-time limit~[52], $\delta^{(t)}$ is replaced by a delta distribution and the (constant) dichroic signal reads
\begin{align}
\Delta \Gamma (\omega) &= 4\pi \mathcal{E}^2 \sum_{\mathbf{k}} \sum_{n>1} \Omega^n_{xy} (\mathbf{k}) \, \delta (E_n(\mathbf{k}) - E_1(\mathbf{k}) - \omega), \notag\\
&=4\pi \mathcal{E}^2 \sum_{\mathbf{k}} \sum_{1<n<q} \Omega^n_{xy} (\mathbf{k})\,  \delta (E_n(\mathbf{k}) - E_1(\mathbf{k}) - \omega),\notag
\end{align}
where we introduced the quantity $\Omega^n_{xy} (\mathbf{k})$ defined in Eq.~\eqref{eq: Berry curvature}, and then used the result $\sum_{\mathbf{k}}\Omega^q_{xy} (\mathbf{k})\!=\!0$ obtained in Eqs.~\eqref{qzero}-\eqref{qodd} together with the symmetry in Eq.~\eqref{energies_relation}.

This demonstrates that the dichroic signal $\Delta \Gamma (\omega)$ has no contribution from the higher band, for any value of the flux. Specifically, this explains the absence of signal in Fig.1c (main text) in the frequency regime $\omega\!>\!4J$.

We point out that the results derived in this Appendix entirely rely on the sublattice (chiral) symmetry of the model [Eq.~\eqref{pi_operator}], and hence, they remain valid for any lattice model that satisfies a similar property.

\subsection{Transverse optical conductivity of a fractional Chern insulator}

\begin{figure}[h!]
\includegraphics[width=0.98\linewidth]{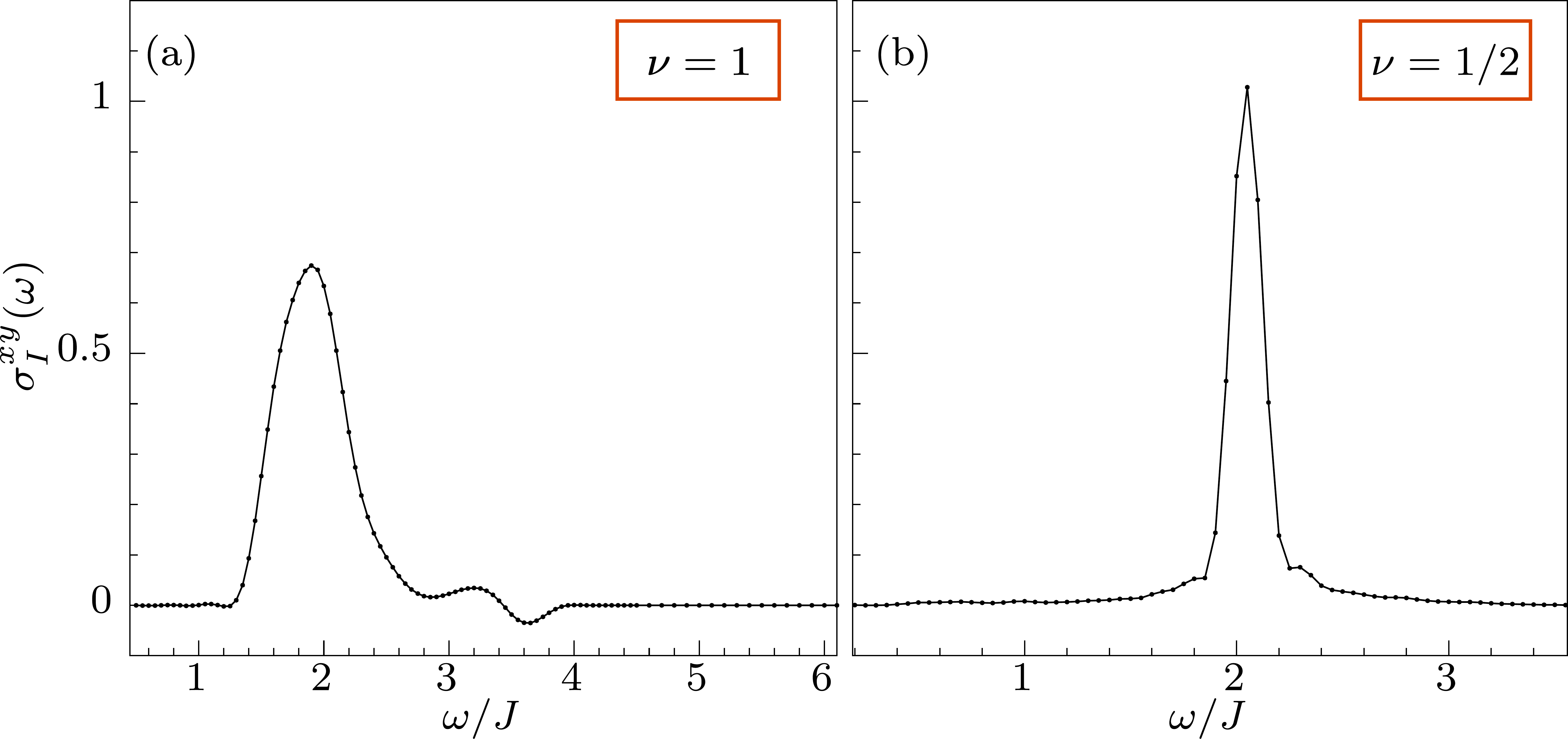}
\caption{Transverse optical conductivity of the Hofstadter model (a) for non-interacting fermions at $\nu = 1$ (Chern insulator) and (b) for hard-core bosons at $\nu = 1/2$ (FCI).}
\label{fig: optical_conductivity}
\end{figure}

In the main text, we have emphasized the significance of the integrated dichroic signal $\Delta \Gamma^{\text{int}}$, as a probe for topological order in strongly-correlated matter. However, the dependency of the dichroic signal $\Delta \Gamma (\omega)$ on the drive frequency $\omega$ is also of interest:~it is proportional to the imaginary part of the antisymmetric optical conductivity~[43, 45, 47]
\begin{equation}
\sigma_I^{xy} (\omega) = \hbar \omega \Delta \Gamma (\omega)/4 A_{\text{syst}}\mathcal{E}^2 . \label{optical_conductivity_suppmat}
\end{equation}
In Fig.~\ref{fig: optical_conductivity}, we show the optical conductivity $\sigma_I^{xy} (\omega)$ associated with the Chern insulator ($\nu\!=\!1$) and the fractional Chern insulator ($\nu=1/2$) realized in our model. We note that these results complement studies of the optical conductivity in the integer quantum Hall effect~[81, 82]. Interestingly, the plot presented in Fig.~\ref{fig: optical_conductivity}b shows that the optical conductivity of the FCI is strongly peaked around the frequency $\omega\!\approx\!2J$, which indicates that this transport coefficient has dominant contributions from inter-band transitions; see also the main text for a similar analysis of the DIR $\Delta \Gamma^{\text{int}}$.

\begin{thebibliography}{78}
\bibitem [{\citenamefont {we~have further studied the entanglement spectrum on
  either side of the CDW-FCI~phase transition}()}]{footnote}%
  \BibitemOpen
  \bibfield  {author} {\bibinfo {author} {\bibnamefont {We have further studied
  the entanglement spectrum on either side of the CDW-FCI~phase transition}},\
  }\href@noop {} {\bibinfo  {journal} {and we have verified that it has the
  features expected for these respective two phases. The variation of the
  entanglement entropy across the phase transition [Fig.4b] reflects these
  characteristic features}\ }\BibitemShut {NoStop}%
\end{thebibliography}
\end{document}